# Complex Network Approach to Human Promoter Sequences

Huijie Yang[1], Fangcui Zhao[2], Binghong Wang[1]

1. Department of Modern Physics, University of Science and Technology of China, Hefei Anhui 230026, China

2. College of Life Science and Bioengineering, Beijing University of Technology, Beijing 100022, China

Abstract

Based upon the correlation matrix of the human promoter sequences, a complex network is constructed to capture the principal relationships between these promoters. It is a complex network has the properties of the right-skewed degree distribution and the clustering simultaneously, i.e., a hierarchical structure. An eigenvector centrality (EC) based method is used to reconstruct this hierarchical structure.

PACS (numbers): 89.75.Hc; 87.14.Gg; 89.15.Fb

As one of the most exciting topics in molecular genetics, gene regulation attracts special attentions in recent years [1]. The quantity of information gained in the sequencing and gene expression projects both requires and enables us to solve this problem [2-4]. Promoter sequences are crucial in gene regulation. The general recognition of promoters and the analysis of these regions to identify the regulatory elements in them are the first step towards complex models of regulatory networks.

Suzuki et al. used the oligocapping method to identify the transcriptional start sites from cDNA libraries enriched in full-length cDNA sequences, which they have made available at the Database of Transcriptional Start Sites (DBTSS; http://dbtss.hgc.jp/) [5]. Consequently, the authors in reference [6] have used the DBTSS data set and aligned the full-length cDNAs to the human genome, thereby extracting putative promoter regions. Using the known transcriptional start sites from over 5700 different human full-length cDNAs, a set of 4737 distinct putative promoter regions (PPRs) are extracted from the human genome. Each PPR consists of nucleotides from -2000 to +1000 bp relative to the corresponding transcriptional start site. Preliminary results, such as the CpG islands, the transcription factor-binding sites, the over- and under-representation of words and the deterministic structures, show that the regularity elements cluster around the transcriptional start site. Herein, we consider the correlations between all the 4737 PPRs from human genome.

The correlations in DNA sequence have been an important topic in bioinformatics [7-19]. The correlation properties of bio-sequences may shed light on the corresponding biological functions. In this paper, from the viewpoint of complex network we investigate the correlations between the Human PPRs available in Ref. [6]. We found the coexistence of clustering and right-skewed degree distribution, which shows that the relations between the promoters behave a hierarchical manner. The procedure suggested in one of our works is used to reconstruct this hierarchical structure.

Denoting the element of the $i'th$ PPR in position $k$ as $X_i(k)$, we can define the distance between a pair of PPRs as,

$$R_{i,j} = \frac{\sum_{k=1}^{3001} D(X_i(k), X_j(k))}{3001}. \qquad (1)$$

where,

$$D(X_i(k), X_j(k)) = \begin{cases} 0 & X_i(k) = X_j(k) \\ 1 & X_i(k) \neq X_j(k) \end{cases}. \qquad (2)$$

The matrix $R$ is called distance matrix. Introducing a criterion $r_{criterion}$, we can convert the distance matrix into a new matrix describing the principal relations between the PPRs. The convert rule reads,

$$T_{i,j} = \begin{cases} 1 & R_{i,j} \leq r_{criterion} \\ 0 & R_{i,j} > r_{criterion} \end{cases} \qquad (3)$$

Considering each PPR as a node, the relations between PPRs form a complex network, which can be represented with the matrix $T$ exactly.

The value of $r_{criterion}$ is essential in this approach. If it is unreasonable large, the characteristics of the relations between the PPRs are submerged in a large amount of random noises. On the contrary, if it is too small, large statistical fluctuations will destroy the characteristics completely. A special dynamical process can describe this change of $r_{criterion}$. That is, the number of edges decreases while the number of nodes keeps unchanged.

The pioneering work of the small world model [20] stimulated an avalanche of works on complex networks. Complex network theory has been a powerful tool in dealing with problems in diverse research fields, especially in biology [21]. The degree distribution and the clustering coefficient are two important measures of the structure characteristics of networks. The degree of a node is the number of its nearest neighbors. The degree distribution function of a complex network reflects the non-homogenous of the importance of the nodes. The clustering characteristic of a node can be represented with the connecting probability between two neighbors of this node. The clustering coefficient of a network is the average of the clustering characteristic over the whole network. Denoting the neighbors of the node $s$ with $(s_1, s_2, \cdots, s_k)$, each pair of

$(s_i, s_j)$ can leads to two triples $s_i \to s \to s_j$ and $s_i \leftarrow s \leftarrow s_j$. If there is an edge between these two neighbors, the three nodes $(s, s_i, s_j)$ can form two loops $s \rightleftarrows s_i \rightleftarrows s_j \rightleftarrows s$, called triangles in this paper. Hence, the clustering coefficient can be defined as,

$$C = \frac{N_\Delta}{N_3} = \frac{\frac{1}{6} \cdot \sum_{i,j,l}^{N} A_{ij} \cdot A_{jl} \cdot A_{li}}{\frac{1}{2} \cdot \sum_{i=1}^{N} k(i) \cdot [k(i)+1]} = \frac{tr(A^3)}{3 \cdot \sum_{i=1}^{N} k(s) \cdot [k(s)+1]} \quad (4)$$

where $N_\Delta$ is the number of the loops containing three nodes, $N_3$ the number of the connected triples and $k(s)$ the degree of the node $s$.

For an Erdos-Renyi network, where each pair of nodes is connected with a certain probability, the degree distribution obeys a Poisson function and the clustering coefficient is very small. Real world networks have completely different characteristics compared with the Erdos-Renyi networks. One characteristic is the small-world effect, that is, some nodes are connected tightly together to form a community, which can induce large clustering coefficient. The other one is the right-skewed degree distribution, i.e., the degree distribution obeys a power-law $p(k) \propto k^{-\gamma}$. These two properties coexist in many real world networks, to cite examples, in the metabolic networks in a cell, some technique collaboration and social relation networks. This coexistence tells us that the communities should be combined together in a hierarchical way [22-27]. The hierarchy is one of the common features investigated in detail recently. In one of our recent works [28] we proposed a procedure to reconstruct the hierarchical structure in a complex network. To keep this paper self-contained we review that procedure briefly.

Consider the network $T$ constructed from the correlation matrix $R$ and the criterion $r_{criterion}$. The eigenvector centrality (EC) [29-31] is employed as the proxy of importance of each node. Denote the eigenvector corresponding to the principal eigenvalue of this adjacency matrix with $e_{principal}$, the eigenvector centrality of the $i$'s node is the $i$'s component of $e_{principal}$, i.e., $EC(i) = N_0 \cdot |e_{principal}(i)|^2$, where $N_0 = 4737$ is the number of the PPRs. This measure simulates a mechanism in which each node affects all of its neighbors simultaneously.

Calculate the EC values for all the PPRs from the original adjacent matrix $A_1 = T$. Introducing a critical EC value, $EC_{crit} = \gamma \cdot \max(EC)$, the PPRs whose EC values are larger than $EC_{crit}$ can be regarded as the key PPR. We can adjust the parameter $\gamma$ in the range of $[0,1]$. Then remove the found key PPRs from the initial adjacent matrix and obtain a new adjacent matrix $A_2$.

Iteration of this step leads to some sets of key PPRs as, $S_m(n) | m = 0,1,2,3,\cdots,M$ and the corresponding adjacent matrix, $A_m | m = 1,2,3,\cdots,M$. Here, $S_m(n)$ is the set containing $n$ key PPRs found at the $m$'th step. Each set of key PPRs can form a backbone in the corresponding group. Catalogue the PPRs in each set into several levels we can obtain the intra-group structure. Generally, we can sort the key PPRs in each group in an ascending way.

Define the average number of edges per key PPR, and the average number of edges per left PPR as,

$$D_{key}(m) = \frac{K_{key}(m)}{N_{key}(m)},$$
$$D_{left}(m) = \frac{K_0 - K_{key}(m)}{N_0 - N_{key}(m)},$$
(1)

where $K_{key(m)}, N_{key}(m), N_0$ and $K_0$ are the edges between the key PPRs found up to the $m$'th step, the total number of the key PPRs found up to the $m$'th step, the total numbers of PPRs and edges in the original network $T$, respectively. The terminate criterion can be designed as $D_{left}^c$. Once the average edges per left PPR $D_{left}(m)$ decreases to $D_{left}(m) < D_{left}^c$, the procedure is terminated.

Fig.1 to Fig.4 present the degree distribution functions for different values of $r_{criterion}$. With the decrease of $r_{criterion}$, the degree distribution function changes from a Poisson to almost a perfect power-law form. The corresponding network tends to be a scale-free network. During this dynamical process, the random noise due to the loose criterion is filtered out effectively and the characteristics in the PPRs' principal relations become much more significant.

For each network with a certain value of $r_{criterion}$ we construct a corresponding Erdos-Renyi network with the same numbers of nodes and the edges. From the clustering coefficients as shown in Fig.5 we can find that the clustering property does not change significantly with the decrease of $r_{criterion}$. The clustering coefficient for the network at $r_{criterion}=0.72$ is much larger than that for the corresponding Erdos-Renyi network. The network at $r_{criterion}=0.705$ has still a much large clustering coefficient while the corresponding Erdos-Renyi network has a perfect tree-like structure.

This coexistence of clustering and right-skewed degree distribution reveals the hierarchy property of the correlations in PPRs. Fig.6 presents the adjacent matrix of the network with $r_{criterion}=0.705$. The nodes numbered 1-100 in the new ordered adjacent matrix form a core, where the nodes connected tightly. The other sub-groups are loosely connected with this core, while they link with each other with few edges. The size of the core is much larger than that of the other sub-groups. The fine structure of the core is also presented in Fig.7. We can find that the core has a similar structure compared with the total adjacent matrix. By this way, we can classify the promoters into several classes. The average edges among the key nodes and the left nodes as shown in Fig.8 tell us that the groups we obtained are connected much tightly than the left part of the network.

In summary, the autocorrelation matrix between the human promoter sequences is converted to an adjacent matrix of a complex network representing the principal correlations between these promoters. This network displays the coexistence of the right-skewed degree distribution and the clustering characteristic. There are several groups whose nodes are connected tightly. And each group contains several sub-groups. The structure of a group and that of the whole network are similar, which tell us that the network has a typical hierarchical structure. This analysis may give us the relations between the promoters.

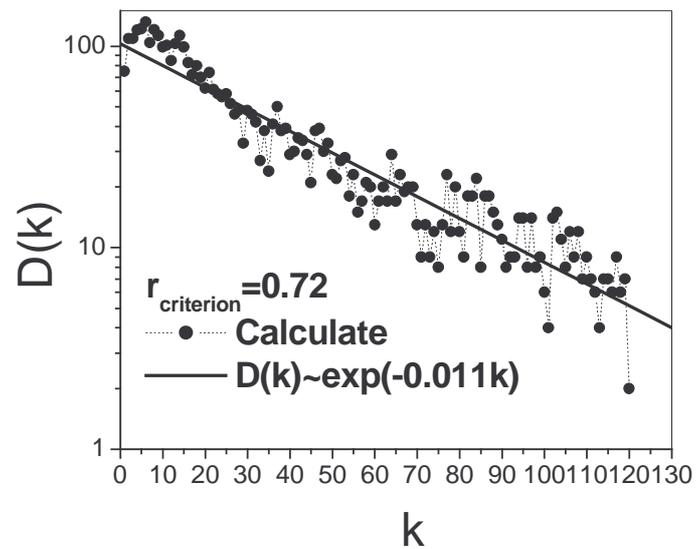

**Fig.1** The degree distribution for the network with the criterion $r_{criterion} = 0.72$. It obeys an almost perfect Poisson-law, $p(k) \propto e^{-0.011k}$.

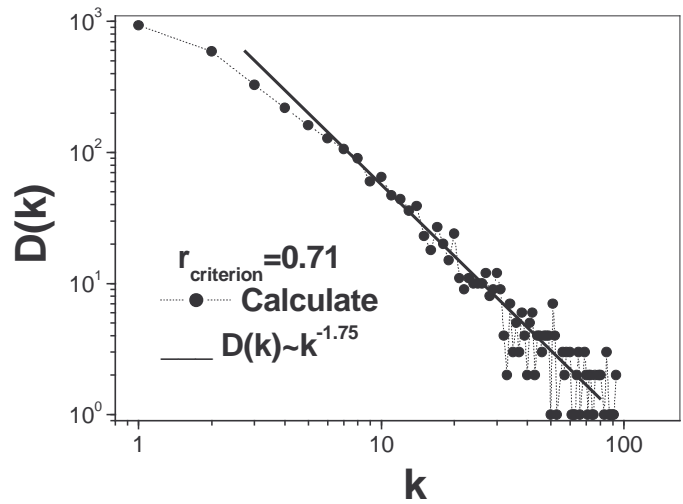

**Fig.2** The degree distribution for the network with the criterion $r_{criterion} = 0.71$. It obeys roughly a power-law, $p(k) \propto k^{-1.75}$. There is a little down-bend at the beginning of the curve, which will be improved with the decrease of $r_{criterion}$ as shown in the Fig.3 and Fig.4.

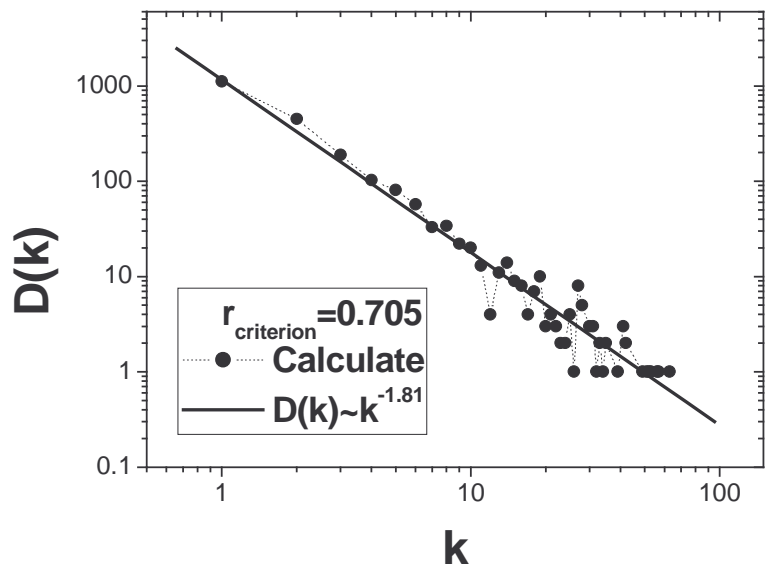

**Fig.3** The degree distribution for the network with the criterion $r_{criterion} = 0.705$. It obeys an almost perfect power-law, $p(k) \propto k^{-1.81}$.

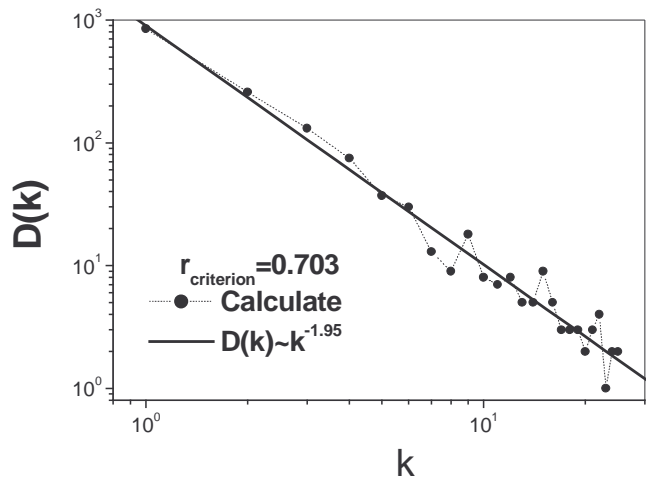

**Fig.4** The degree distribution for the network with the criterion $r_{criterion} = 0.703$. It obeys an almost perfect power-law, $p(k) \propto k^{-1.86}$.

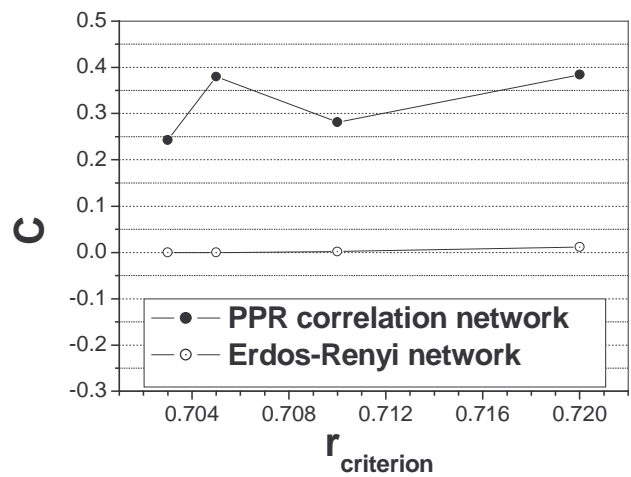

**Fig.5** The clustering coefficients for the PPR correlation networks and the constructed corresponding Erdos-Renyi networks. The clustering property does not change significantly with the decrease of $r_{criterion}$. The clustering coefficient for the network at $r_{criterion} = 0.72$ is much larger than that for the corresponding Erdos-Renyi network. The network at $r_{criterion} = 0.705$ has still a much large clustering coefficient while the corresponding Erdos-Renyi network has a perfect tree-like structure.

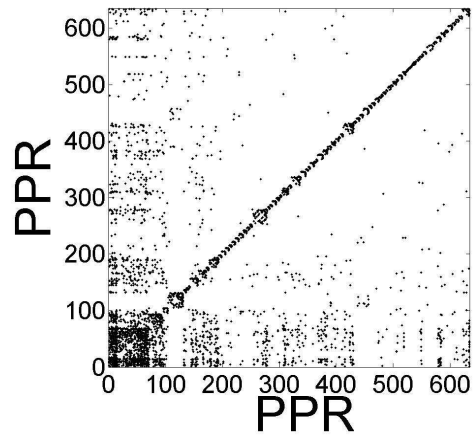

**Fig.6** The new-ordered adjacent matrix for the PPR correlation network with the critical value $r_{criterion} = 0.705$. The parameters $\gamma = 0.1$ and $D_{left}^{c} = 1.0$. The nodes numbered 1 to 100 in this adjacent matrix form a core, where the nodes connected tightly. The other sub-groups are loosely connected with this core, while they link with each other with few edges. The size of the core is much larger than that of the other sub-groups.

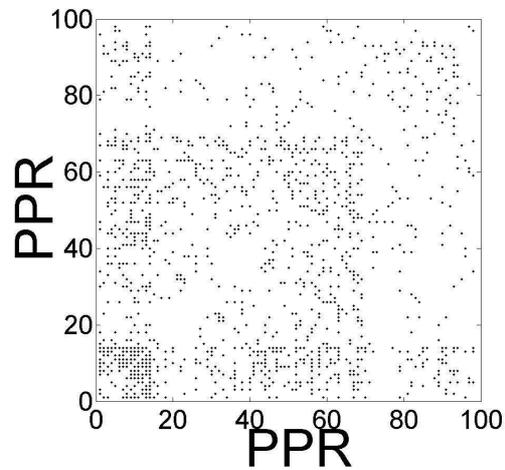

**Fig.7** Enlargement of the largest subgroup (core) in Fig.6. The core has a similar structure compared with the total adjacent matrix. By this way, we can classify the promoters into several classes.

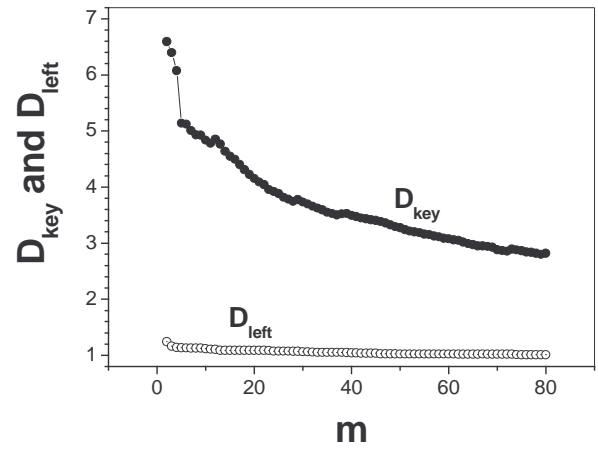

**Fig.8** The average edges among the key nodes and the left nodes. We can find that the groups we obtained are connected much tightly than the left part of the network.